\documentstyle[preprint,aps,prl]{revtex}

\begin{document}
\draft

\title{ Large scale quantum simulations:
C$_{\bf 60}$ impacts on a semiconducting surface
}

\author{Giulia Galli and Francesco Mauri}

\address{Institut Romand de Recherche Num\'{e}rique
en Physique des Mat\'{e}riaux (IRRMA),\\
INR-Ecublens, CH-1015 Lausanne, Switzerland}
%
% \receipt{}
%
\maketitle
\begin{abstract}
 We present tight binding molecular dynamics
simulations of  C$_{\rm 60}$ collisions on the reconstructed
diamond(111) surface, carried out with an O(N) method and with cells
containing 1140 atoms.
The results of our simulations are in very good agreement with
experiments performed under the same impact conditions.
Furthermore our calculations provide a detailed characterization of
the microscopic processes occuring during the collision,
and allow the identification of three impact regimes,
as a function of the fullerene incident energy.
Finally, the study of the reactivity between the cluster and
the surface gives insight into the deposition mechanisms of
C$_{\rm 60}$ on semiconducting substrates.

\end{abstract}
\newpage
\narrowtext

 Since the discovery\cite{BJADW91} of the unique stability of
C$_{\rm 60}$ molecules against surface induced fragmentation,
the investigation of fullerene interactions with solid substrates
has become an exciting field of research$^{2-8}$.
 Collisions of C$_{\rm 60}$ on several surfaces
have been reported to induce no fragmentation of the impinging molecules, for
initial kinetic energies (E$_{\rm k}$)
up to a few hundred eV\cite{BJADW91}.
This property is unprecedented in
molecular-ion surface-induced phenomena and has earned
C$_{\rm 60}$ the name of {\it resilient} molecule.

 Of particular interest is the case of C$_{\rm 60}$ impacts
on semiconducting surfaces which can form covalent bonds with the
cluster. The study of these bonds
is a challenging problem, from a fundamental point of view. Furthermore
such an investigation is an essential prerequisite for the
understanding of fullerene deposition on non-metallic substrates.
These deposition processes have recently been proposed\cite{PMDPPC93}
as a valuable way of synthesizing thin films which would retain
specific characteristics of the incident clusters.

 In order to understand the physics of C$_{\rm 60}$ collisions
on solid substrates, it is crucial to reveal the processes
occurring {\it at} the surface.
Whereas these processes are difficult to probe experimentally\cite{BJADW91},
they can be investigated by computer simulations.
Recently, simulations of C$_{\rm 60}$ impacts on surfaces
(hydrogenated diamond and graphite) have been performed by using empirical
classical potentials \cite{MBDMW91}.
These calculations constitute a useful starting point for the study
of fullerenes on surfaces. However, a more accurate description
of the cluster-surface interactions is desirable.
In principle this can be obtained by molecular dynamics
(MD) simulations in which the interatomic forces are derived
at each step from quantum mechanical (QM) computations.
In practice, realistic simulations of C$_{\rm 60}$ interactions with a surface
require cells with at least several hundreds of atoms.
Until recently QM simulations of systems of this size have not been
possible because of the computer time required
by conventional algorithms, which grows as the cube of the number of atoms
in the system (N). Indeed the only semiempirical QM study of
C$_{\rm 60}$ collisions with a surface (hydrogenated diamond)
reported to date \cite{BFBL94} has been limited to small MD cells.

 This situation has changed with the introduction of
methods\cite{MGC93,MG94,LNV93}
for electronic structure calculations and MD simulations, which are
based on algorithms whose computational workload grows linearly
with the system size.
These approaches  (often referred to as O(N) methods) have opened the way
to QM-MD simulations of systems much bigger
than previously accessible,
and therefore to the study of problems usually out of reach
for QM calculations.

 In this Letter we present QM-MD simulations of  C$_{\rm 60}$ impacts
on a semiconducting surface as a function of the molecule incident energy,
carried out with an O(N) method \cite{MG94}, and with MD cells
containing 1140 atoms.
Interatomic forces are described according to a
tight binding (TB) Hamiltonian\cite{xwch92},
derived from first principles calculations.
We considered the clean (2 $\times$ 1)
reconstructed diamond (111) surface.
This surface can be expected to form bonds with
the impinging C$_{\rm 60}$, since its uppermost layer contains
three fold coordinated $\pi$ bonded atoms; it therefore represents
a good candidate for the study of covalent bonds between the fullerene and
a semiconducting substrate.
Our work constitutes the first QM study of such interactions and
the first large scale TB-MD simulation based on an O(N) methodology.
 The results of our calculations are in very good agreement
with experiments carried out under the same impact conditions
\cite{BJADW91,BLH91,BLRH92,LBRH92,YW92}. Furthermore they allow us to
characterize in detail the microscopic processes occuring during the collision
and thus to identify three different impact regimes as a function of
C$_{\rm 60}$ incident energy.

 In our calculations, we used the O(N) method of Ref.~\cite{MG94}.
This approach is based on an energy functional with implicit orthogonalization
constraints and on a localized orbital (LO)
formulation. We adopted the TB Hamiltonian (${\cal H}$)
proposed by Xu et al.~\cite{xwch92,ftn1}.
The LOs were centered on atomic sites, extending up to second neighbors
(in the notation of Ref.~\cite{MG94}, N$_{\rm h}$ = 2)\cite{ftn2}.
 We simulated neutral C$_{\rm 60}$ molecules colliding
with a reconstructed C(111) surface at initial energies
E$_{\rm k}$= 60, 80, 120, 150, 180, 210, 240, 300, 400 eV.
Each of the nine runs lasted from 0.3 to 1 ps.
The efficiency of the O(N) algorithm allowed us to carry out all
calculations on workstations.
C$_{\rm 60}$ impinged upon the surface at normal incidence,
oriented in such a way that a double bond between two hexagons faced
a Pandey chain of the surface.
The C(111) substrate was represented by a slab
composed of twelve layers, each containing ninety C atoms.
The dimensions of a layer were 22.72$\times$ 21.86 \AA\ in the
$x$ and $y$ direction, respectively.
The slab was terminated on each side by a reconstructed surface, and
periodic boundary conditions were applied along $x$ and $y$.
 Nine surface layers were allowed to move during the simulations.
According to the size of our MD cell and to the number of mobile
layers, we estimate that the shock wave produced by a collision of
C$_{\rm 60}$ with the substrate is echoed  back to
the impact region about 0.2 ps after the impact.
This is the time interval during which we observe the formation
of bonds between the cluster and the surface.
Therefore the characterization of the three impact energy regimes (ERs)
proposed in the following is not
affected by finite size effects.

 Before starting the simulation of  C$_{\rm 60}$ collisions on the surface,
we optimized the slab and the molecule geometries independently.
Our results for the surface reconstruction
compare well both with those of a conventional (extended orbital[EO])
calculation using the same ${\cal H}$\cite{DP94},
and with the findings of first principles computations\cite{IGGPT92}.
For example, we find that the average bond distance in the surface chains
is 1.44 \AA\  and that the change in bond length
between the second and third layer is 6$\%$; the corresponding values
reported in Ref.~\cite{DP94} are 1.48 \AA\ and 6$\%$, whereas those
of Ref.~\cite{IGGPT92} are 1.44 \AA\ and 8$\%$.
However our calculation seems to overestimate the
dimerization of the surface chains (4$\%$ versus 0.2$\%$
in Ref.~\cite{DP94} and 1.4$\%$ in Ref.~\cite{IGGPT92}).
The two characteristic bond lengths of C$_{\rm 60}$ calculated within the
LO formulation  differed by $5\%$ (single bond) and $2.5\%$
(double bond) from those optimized with EO\cite{ftn22}.
Also in this case the TB results\cite{Iowac60} are in good agreement
with those of first principles calculations\cite{FAPC91}.

 We now turn to the discussion of the different regimes
of C$_{\rm 60}$-surface interactions which we observed in our simulations.
For E$_{\rm k}$=60, 80 and 120 eV (low ER)
the molecule does not form any bond with the surface\cite{ftn3}.
The minimum distance between C$_{\rm 60}$ atoms and the
substrate ({\bf D})
is approximately constant during the collision
and very close to 2 \AA, as seen in Fig.~1, where we show
{\bf D} as a function of the simulation time ($t$).
The impact provokes large distortions in the fullerene cage:
the cluster height is decreased from 7.0 to 3.9, 3.4  and 3.1\AA\ ,
for E$_{\rm k}$= 60, 80 and 120 eV,
respectively, and the flattened molecule adapts its shape to
that of the reconstructed surface (see Fig.~2).
For E$_{\rm k}$=120 eV, a few bonds of the molecule are broken
during the collision, whereas no bond breaking is observed
at lower energies. This can be seen in
Fig.~3 which displays the number of differently coordinated sites of
C$_{\rm 60}$ as a function of $t$.
In all of the three cases,
after the impact the cluster
leaves the surface with its original shape, without any defect.
The kinetic energies of the centre of mass (E$_{\rm cm}$)
of the outgoing molecule is shown in Fig.~4.
In this ER, E$_{\rm cm}$ shows a clear dependence
upon E$_{\rm k}$. In particular E$_{\rm cm}$ is proportional to E$_{\rm k}$ for
incident energies  up to 80 eV, i.e. in the energy interval in which
no bonds of the cluster are broken.

 For  E$_{\rm k}$ larger than 120 and smaller than 240 eV
(medium ER), the molecule does not only suffer large
distortions upon impact on the surface, but also forms bonds
with the substrate. During the collision, some bonds of the
C$_{\rm 60}$ cage are broken, resulting in the formation of twofold
coordinated sites.
The presence of defects in the C$_{\rm 60}$ structure causes the molecule
to react with the surface, at variance with what happens in the low ER.
Indeed, after some oscillations
around 2 \AA, {\bf D} suddenly
decreases to a value of about 1.5 \AA, i.e.
to a distance typical of sp$^3$ like bonds (see Fig.~1).
Not surprisingly the surface atoms involved
in the bonding belong to the topmost layer,
containing three-fold coordinated sites in the Pandey reconstruction.
In this ER, the number of bonds between C$_{\rm 60}$ and the surface
depends weakly on E$_{\rm k}$ and, during the collision, is approximately two.
For E$_{\rm k}$=150 eV, we find that these bonds are stable
after the impact and that the cluster is adsorbed on the surface
(see Fig.~2). We note that a decrease in scattering intensities
around 140 eV  has been found in collision experiments on a
graphite substrate\cite{BLH91,YW92}, and that adsorption of intact C$_{\rm 60}$
was invoked by Busman et al.\cite{BLH91} in order to explain this observation.
At E$_{\rm k}$=180 and 210 eV, the C$_{\rm 60}$-surface
bonds break after the collision and the molecule leaves
the substrate with E$_{\rm cm}$ = 8 and 15 eV, respectively.
Our findings compare well with those of Ref.~\cite{BLH91},
where the E$_{\rm cm}$ of fullerene ions scattered from graphite
is found to be about 10 to 20 eV and nearly independent of
the impact energy, for E$_{\rm k}$ larger than 140 eV.
%\cite{BJADW91,BLH91,BLRH92,LBRH92}.
In this impact regime, after the collision,
the topology of the molecule is still
that of a cage, although defects are present with respect to the
original shape (see Figs.~2 and 3).

 Upon impact at E$_{\rm k}\geq 240$ eV (high ER),
we observe the formation of several bonds between C$_{\rm 60}$
and the surface. Their average number increases with the incident energy,
varying from about six to sixteen.
Shortly after the collision, many bonds break within the
molecule and the cage structure
of the fullerene can no longer be identified (see Fig.~4);
a disordered structure is formed and adsorbed on the substrate.
After the 300 and 400 eV collisions,  some of the C$_{\rm 60}$ atoms form
twofold coordinated chain structures (see Fig.~2); the formation of such
chains was observed also in MD studies of fullerene melting \cite{KT94}.

 Whenever C$_{\rm 60}$ left the surface,
we did not observe any fragmentation of the molecule.
However a rupture may be expected to occur some time after
a collision if the internal energy (E$_{\rm in}$) of the bouncing molecule
(see Fig.~4) exceeds its stability threshold.
MD investigations of fullerene melting \cite{KT94} have shown
that  C$_{\rm 60}$ becomes unstable when its internal kinetic
energy (E$_{\rm in}$) is between 30 and 40 eV.

 In the low and medium ER the topology of the reconstructed surface
is unaffected by the collision, although a considerable fraction of
E$_{\rm in}$ is transferred from the molecule to the surface
(e.g. as much as 50$\%$ for E$_{\rm k}$ = 180 and 210 eV).
In the medium ER only the three fold coordinated atoms of
the topmost layer bond to C$_{\rm 60}$ atoms (see Fig.~2).
In the high ER the fourfold coordinated surface atoms
(belonging to the second layer) are also involved
in bonding to the molecule. However they remain fourfold
coordinated since their bonds with the third layer atoms are broken.
This bond breaking induces a local surface dereconstruction
from a $\pi$ bonded chain geometry towards a (1 $\times$ 1) ideal
arrangement (see Fig.~2).

 We finally note that the connection between
C$_{\rm 60}$ enhanced reactivity with the surface and the presence
of defects in the molecule might characterize deposition
processes of the fullerene on other semiconducting surfaces.
Therefore the variety of STM images produced by
covalently bonded C$_{\rm 60}$ on, e.g., Si(111) \cite{LCPWCS92},
might correspond to molecules with different types of defects.

 In summary, we have identified three impact energy regimes of
C$_{\rm 60}$ collisions on C(111).
For E$_{\rm in} \leq 120$ eV, the molecule bounces off the surface
without ever forming bonds, and recovers its original shape after
severe distortions.  In a second regime, C$_{\rm 60}$ forms covalent
bonds with the surface. The cluster can be either adsorbed on
or leave the substrate, with defects
in its original structure.
The energy at which we observe
absorption (150 eV) corresponds to the energy at which
a decrease of fullerene ions scattered
from graphite is observed experimentally.
In the third regime
(E$_{\rm k} \geq 240$ eV ) the molecule
breaks after the impact and pieces of the broken cage form
stable bonds with the substrate, which can induce a local
surface dereconstruction. The formation of bonds
between C(111) and the fullerene is always
accompanied by the formation of defects in the molecule.
When the molecule bounces off the surface, the kinetic energy
of its centre of mass shows a clear dependence on its incident
energy in the low ER; on the contrary
E$_{\rm cm}$ depends weakly on E$_{\rm in}$ in the medium ER
and acquires values in agreement with those measured under
the same impact conditions.
Finally we have shown that QM simulations
for systems containing thousands of atoms, so far not possible with
conventional methods, are now feasible by using O(N) approaches.
In the simulations reported in this work,
the gain in computer time with respect to conventional
O(N$^3$) methods is estimated to be of the order of 1000.

 We would like to thank R.~Car  and F.~Gygi for useful discussions and
A.~Canning and F.~Gygi for a critical reading of the manuscript.
Part of the calculations were performed on the workstation cluster
at CSCS in Manno.
This work was supported by the Swiss National
Science Foundation under grant No 20-39528.93.

\newpage
\begin{figure}
\caption
{
Minimum distance ({\bf D}) between C$_{\rm 60}$ and the surface as a function
of the simulation time:
solid, long dashed, dashed and dotted lines
correspond to E$_{\rm k}$= 120, 150,
180 and 210 eV, respectively.
}
\end{figure}
\begin{figure}
  \caption
{
Snapshots of the uppermost five layers of the slab and of  C$_{\rm 60}$
for different E$_{\rm k}$ and
at given simulation times ($t$): E$_{\rm k}$ = 120 eV and $t$=0.14 ps
(upper panel); E$_{\rm k}$ = 150 eV and $t$=1.00 ps (middle panel);
E$_{\rm k}$ = 300 eV and $t$=0.33 ps (lower panel). (See text).
Red and green spheres denote threefold and fourfold coordinated atoms
of the slab and  black spheres indicate C$_{\rm 60}$ atoms.
}
\end{figure}
\begin{figure}
\caption
{
Twofold and fourfold coordinated C$_{\rm 60}$ atoms ({\bf N$_{\bf at}$})
as a function of the simulation time for selected E$_{\rm k}$:
dotted, solid and dashed lines correspond to E$_{\rm k}$ = 120, 180 and 300 eV,
respectively.
}
\end{figure}
\begin{figure}
\caption
{
Final energies (E$_{\rm f}$) of C$_{\rm 60}$ bouncing on the
surface as a function of E$_{\rm k}$:
diamonds and crosses indicate the kinetic energy of the centre of mass
and the internal energy of the molecule, respectively.
}
\end{figure}

\begin{thebibliography}{10}
\bibitem{ftn1}
 We used a cutoff radius of 2.30 \AA\ for the hopping parameters of
${\cal H}$ and for the two body repulsive potential, and
U=8 eV for the Hubbard like term (see Ref.~\cite{xwch92}).
In simulations performed with U=0 we observed unphysical charge
transfers, as large as one electron on some sites.
\bibitem{ftn2}
At each MD step ${\cal H}$ was diagonalized iteratively
with a conjugate gradient procedure, by performing  15 iterations per ionic
move and by extrapolating linearly the input wavefunction from the two
previous ionic steps.
The time steps used for the integration of the
equations of motion varied from 15 to 6 a.u. according
to  E$_{\rm k}$.
\bibitem{ftn3}
Throughout the paper we use a cutoff distance of 1.9 \AA\ to define
bonds and atomic coordination.
\bibitem{ftn22}
The differences between the LO and the EO
formulations can be systematically reduced by
using LOs which extend up to third neighbors (N$_{h}$=3).
For example in the case of C$_{60}$ with N$_{h}$=3 the errors  are
$1.1\%$ (single bond) and $0.7\%$ (double bond).
We believe that calculations with N$_{h}$=2
are sufficiently accurate for the purpose of the present investigation.
\end{thebibliography}
\end{document}